\documentclass[aps,prd,onecolumn,showpacs,preprintnumbers,groupedaddress,nofootinbib]{revtex4}
\usepackage{graphics}  
\usepackage{epsfig}
 \usepackage{verbatim}

\IfFileExists{srcltx.sty}{\usepackage[active]{srcltx}}

\newcommand{\be}{\begin{equation}}
\newcommand{\ee}{\end{equation}}
\newcommand{\bea}{\begin{eqnarray}}
\newcommand{\eea}{\end{eqnarray}}

\begin{document}

\title{The ground states of baryoleptonic Q-balls in supersymmetric models}

\author{Ian M. Shoemaker}
\author{Alexander Kusenko}
\affiliation{Department of Physics and Astronomy, University of California, Los 
Angeles, CA 90095-1547, USA }

\preprint{UCLA/08/TEP/24}

\begin{abstract}
In supersymmetric generalizations of the Standard Model, all stable Q-balls are associated with 
some flat directions.  We show that, if the flat direction has both the baryon number and the lepton
number,  the scalar field inside the Q-ball can deviate slightly from the flat direction in the
 ground state. We identify the true ground states of such nontopological solitons, including
the electrically neutral and electrically charged Q-balls.
\end{abstract}

\maketitle

\section{Introduction}

Nontopological solitons, Q-balls appear in theories with some scalar fields carrying a global
conserved quantum number when the scalar potential meets the conditions for
Q-ball stability with respect to a decay into scalar
particles~\cite{Friedberg:1976me,Coleman:1985ki,Lee:1991ax}.  These conditions are 
satisfied in supersymmetric generalizations of the Standard Model for the scalar fields carrying the
baryon and lepton numbers~\cite{Kusenko:1997zq}.   In theories 
with gauge-mediated supersymmetry breaking, Q-balls can be stable with respect to decay into both scalar particles and fermions if the vacuum expectation value (VEV) inside the Q-ball corresponds to a flat direction~\cite{Dvali:1997qv}.  These stable objects can form in the early universe from the fragmentation of the Affleck--Dine~\cite{Affleck:1984fy,Enqvist:2003gh,Dine:2003ax} condensate,
and they can constitute all or part of cosmological dark matter~\cite{Kusenko:1997si}.   The role of
supersymmetric Q-balls in cosmology and astrophysics has been the subject of many 
studies~\cite{Frieman:1988ut,Frieman:1989bx,Griest:1989bq,Griest:1989cb,Enqvist:1997si,
Enqvist:1998ds,Enqvist:1998xd,Kusenko:1997it,Kusenko:1997vp,Kusenko:1997hj,Laine:1998rg,
Enqvist:1998en,Enqvist:1998pf,Axenides:1999hs,Banerjee:2000mb,Battye:2000qj,Allahverdi:2002vy,
Kusenko:2004yw,Kusenko:2005du,Berkooz:2005rn, Berkooz:2005sf,Kusenko:2008zm,
Johnson:2008se,Kasuya:2008xp,Sakai:2007ft,Campanelli:2007um,Kasuya:2000wx,
Kawasaki:2005xc,Kasuya:2007cy}.  However,  it was always assumed that the scalar field that makes
up the Q-ball has its VEV aligned exactly along the flat direction.  In
this paper we will show that, for a class of flat directions, this is not the case, and we will
describe the true ground states of the corresponding Q-balls.

What makes the Q-ball stable is a  combination of the baryon number conservation and the energy
conservation.  By construction, a Q-ball has a lower mass than a collection of the scalar particles
carrying the same baryon number.   However, one should also consider the possibility of a decay
into fermions~\cite{Cohen:1986ct}.  Very large Q-balls are likely to form from
the fragmentation of the AD
condensate~\cite{Kasuya:1999wu,Kasuya:2000sc,Kasuya:2000wx,Kusenko:1997si}. For such large
Q-balls with the baryon number $Q_B$, the mass per unit baryon 
number decreases with $Q_B$ as $Q_B^{-1/4}$. When the mass per unit baryon number is below $m_p$,
the proton mass, the Q-ball has a lower mass than a collection of protons and neutrons with the same
baryon number.  Such a Q-ball cannot decay.  The same reasoning would not apply to a purely leptonic
Q-ball, because the lightest fermions carrying the lepton number, neutrinos, have very small masses,
one of which can be arbitrarily close to zero. 

However, in supersymmetric models the scalar fields can carry both the baryon number and the
lepton number. Many flat directions that admit Q-ball solutions have both the baryon and lepton
components related to each other by the requirement that the so called D-terms in the potential
vanish.  All the previous analyses of supersymmetric Q-balls focused on the baryonic squark
component, assuming that the slepton fields just follow the squarks.  This assumption is
reasonable, as long as one considers  general features of the stable Q-balls.   There is an
energetic penalty for deviations of the leptonic and baryonic components.  However, this energetic
penalty may not be sufficient to prevent the emission of a limited number of neutrinos and electrons
from a large baryoleptonic Q-ball.  Such emission can alter the ground state of the relic Q-balls. 
In particular, it can give the electric charge to an otherwise neutral Q-ball.\footnote{Electric
neutrality is a necessary condition for the cancellation of the D-terms.}   Even one unit of
electric charge makes a dramatic difference in the experimental signatures of the relic 
Q-balls~\cite{Kusenko:1997vp}.   For example,  the neutral
Q-balls can be detected by Super-Kamiokande~\cite{Kusenko:1997vp,Arafune:2000yv,Takenaga:2006nr},
while this detector is not well suited for detection of the charged Q-balls.  On
the other hand, the charged Q-balls could be detected by such detectors 
as MACRO~\cite{Kusenko:1997vp,Ambrosio:1999gj} in some range of parameters. The astrophysical limits
on the relic Q-balls may depend on the
electric charge~\cite{Kusenko:1997it,Kusenko:2004yw,Kusenko:2005du}.  
It is, therefore, important to understand the ground states of Q-balls for the purposes of their
detection~\cite{Kusenko:1997vp,Belolaptikov:1998mn,Ambrosio:1999gj,Arafune:2000yv,
Takenaga:2006nr,Cecchini:2008su}.

The paper is organized as follows. In section ~\ref{flat} we review the basic properties of
flat-direction Q-balls in the MSSM. In section~\ref{emission} we consider the emission of neutrinos
from a flat direction Q-ball associated with the $QL \bar{d}$ flat direction. In
section~\ref{neutral} we include the contributions of higher-dimensional operators. In
section~\ref{charged} we calculate the emission of electrons that is energetically allowed. In
section~\ref{both} we show that the relative rates of decay for neutrino and electron emission can
favor a positively charged Q-ball.

\section{Flat-direction Q-balls}
\label{flat}

Minimal Supersymmetric Standard Model (MSSM) has a large number of flat
directions~\cite{Gherghetta:1995dv}. 
Let us begin by considering a toy model that will capture the main features of Q-balls in the MSSM. 

A flat direction in general can be parameterized by a single scalar degree of freedom, although for
many flat directions the relation between the gauge eigenstates and the flat direction parameter
can be non-trivial~\cite{Enqvist:2003pb}.  A Q-ball can form along a flat direction parameterized by
a squark field $q$ and a slepton field $L$. A Q-ball is a nontopological soliton which minimizes the
energy under the constraint of a fixed global
charge~\cite{Friedberg:1976me,Coleman:1985ki,Lee:1991ax}.  Thus
we minimize 
\be E = \int d^{3}x \left[\frac{1}{2} \dot{q}^{2} + \frac{1}{2} |\nabla q|^{2} + \frac{1}{2}
\dot{L}^{2} + \frac{1}{2} |\nabla L|^{2} + U(q,L) \right], \ee
while keeping the the baryonic and the leptonic charges separately conserved:
\be \label{charges} Q_{B}  \equiv \frac{1}{2i} \int d^{3} x \left(q \partial_{0} q^{*} - q^{*} \partial_{0} q \right) \ee
 \be Q_{L}  \equiv \frac{1}{2i} \int d^{3} x \left(L \partial_{0} L^{*} -L^{*} \partial_{0} L \right). \ee

One can solve to the above extremization problem by generalizing the method of Lagrange
multipliers as used in Ref.~\cite{Kusenko:1997ad} to the case of multiple conserved quantum numbers
$Q_a$. One looks for a minimum of 
 \be \label{energy} 
\mathcal{E}_{\omega_{L}, \omega_{q}} =E +\sum_{a =L,B} \omega_{a} \left[ Q_{a} -
\frac{1}{2i} \int d^{3}x \left(a\partial_{0} a^{*} - a^{*} \partial_{0} a \right) \right]  \ee

We use the same approach as in the single field case~\cite{Kusenko:1997ad} to derive the
time-dependence of the fields which comprise the Q-ball. Namely, we exploit the fact that we can
rewrite eq. (\ref{energy}) as
\bea  \label{lagrange} \mathcal{E}_{\omega_{L}, \omega_{q}} &=& \int d^{3}x \left[ \frac{1}{2}
|\dot{L} - i \omega_{L} L|^{2}  +  \frac{1}{2} |\dot{q} - i \omega_{q} q|^{2}  +\frac{1}{2} |\nabla
q|^{2} \right. \nonumber \\   
&&+  \left. \frac{1}{2} |\nabla L|^{2} + \hat{U}(q,L) \right] +  \omega_{L} Q_{L} + \omega_{q} Q_{B}
, \eea
 where $\hat{U} \equiv U(q,L) - (1/2) \omega_{L}^{2} L^{2} - (1/2) \omega_{q}^{2} q^{2}$. Since the
first two terms are non-negative definite and no other term has an explicit time dependence, one can
set them to zero by choosing 
\bea L(r,t) &=& L(r) e^{i\omega_{L} t}, \nonumber \\
q(r,t) &=& q(r) e^{i\omega_{q} t}.  \eea
For definiteness, let us now consider the flat direction for which $L =
q$, with the potential given by $U(L,q) = U_{0} + \lambda |L^{2} - q^{2}|^{2}$, where $U_{0}$ is a
constant.  The value of $\lambda$ has no effect on the Q-ball solution with $\omega_{L} =
\omega_{q} \equiv \omega$ along the flat direction, which, in the thick-wall
regime~\cite{Kusenko:1997ad},  can be approximated as follows: 
\bea \label{here} 
L(r) &=& \left \{ 
\begin{array}{l}
L_{0} \sin (\omega r)/(\omega r) , r \le R \\
L_1 \exp\{-m_L r\}, r>R 
\end{array} \right. \nonumber \\
q(r) &=& \left \{ 
\begin{array}{l}
q_{0} \sin (\omega r)/(\omega r) , r \le R \\
q_1 \exp\{-m_q r\}, r>R 
\end{array} \right. .
\eea
Here the constants $ L_1, q_1$ and $R$ are chosen so that they minimize $\mathcal E_\omega$, while
the solutions are continuous at $r=R$.  We have set the radii of the leptonic and the baryonic
components equal to each other, which results in the lower value of  $\mathcal E_\omega$. 
 For the flat direction  $L_{0} = q_{0}$.   The exponential tails in eq.~(\ref{here}) give a
negligible  contribution to most of the quantities we compute below; in most cases one can assume
that the solution vanishes at $r=R$.  We will use this approximation in what follows.  

Using the profiles of eq.~(\ref{here}) in the expression for the leptonic and baryonic charges 
(\ref{charges}), we get, approximately, 
\bea \label{charges1} Q_{L} = \frac{2 \pi^{2}}{\omega^{2}} |L_{0}|^{2}, \\
Q_{B} = \frac{2 \pi^{2}}{\omega^{2}} |q_{0}|^{2}, \eea
which must be equal by the $L_{0} = q_{0}$ constraint. 

The mass of the Q-ball and the frequency  $\omega$ can be found from the minimization procedure
described above:
\be \label{mass} M(Q_{L},Q_{B} ) = \frac{ 4 \sqrt{2} \pi}{3} U_{0}^{1/4} (Q_{L}+Q_{B})
^{3/4}, \ee
\be \omega(Q_{L},Q_{B})  = \sqrt{2} \pi U_{0}^{1/4} (Q_{L}+Q_{B})^{-1/4}. \ee
One can also estimate the radius $R$ of the flat-direction Q-ball by neglecting the
exponential tail in eq.~(\ref{here}) and setting $L(R) \approx 0$. This yields 
\be \label {radius} R(Q_{L},Q_{B}) = \frac{\pi}{\omega(Q_{L},Q_{B})} = \frac{1}{\sqrt{2}}
U_{0}^{-1/4} (Q_{L}+Q_{B}) ^{1/4}. \ee
eqns. (\ref{mass})-(\ref{radius}) summarize the general features of the Q-ball formed along a flat
direction. 

	\section{Emission of neutral leptons} 
\label{emission}

As discussed in the introduction, the flat-direction Q-balls in MSSM are stabilized by the baryon
number and energy conservation, but an emission of some small number of leptons is an
interesting possibility.  Let us now consider a flat-direction Q-ball carrying some lepton number
and some baryon number, and decaying into an off-flat-direction Q-ball plus some number of fermionic
leptons. In this subsection we consider only the neutrino emission, so that we can ignore Coulomb
effects. We label all initial Q-ball values (the FD values) with an $i$ and all final state Q-balls
with an $f$. 

We assume that the emission of fermions is a small perturbation from the flat direction
case, so that the ansatz $\phi(r) = \sin \omega r /(\omega r)$ for both squarks and sleptons still
holds.  Although away from the direction, the value of mass per unit charge in the condensate
$\omega$ is different from its flat-direction value, the equality of
squark and slepton $\omega$'s still holds for energy reasons. This can be seen by setting
$\omega_{q} \equiv \omega$, $\omega_{L} \equiv \omega + \delta \omega$ in eq.~(\ref{lagrange}) and
minimizing with respect to $\delta \omega$. Then the energy is 
\bea \mathcal{E}_{\omega,\delta \omega} &=& (\omega + \delta \omega)Q_{L} + \omega Q_{B} \nonumber 
\\
&+&  \int d^{3}x \left[ \frac{1}{2}|\nabla q|^{2} + \frac{1}{2}|\nabla q|^{2} + \hat{U}(q,L)\right] 
\eea	
We note that the gradient term for each field cancels exactly with the quadratic term hidden in
$\hat{U}(q,L)$, so that 
\be \mathcal{E}_{\omega, \delta \omega} = (\omega + \delta \omega)Q_{L} + \omega Q_{B}  +
\lambda \int d^{3}x |L^{2} - q^{2}|^{2} .
\ee
The integral in the above expression is equal 
	\be I(\delta \omega) = \frac{\lambda}{\pi^{3}} \int_{0}^{\pi/\omega} \left[q^{2}
\left(\frac{\sin \omega r}{\omega r}\right)^{2} - L^{2} \left(\frac{\sin (\omega+\delta \omega) r}{(
\omega + \delta\omega) r}\right)^{2} \right]^{2} r^{2} dr . \ee 
We use the definition of the U(1) charges in eq.~(\ref{charges}) and take $Q_{L}
\approx Q_{B}$ in the off-flat-direction state to write  
\be I(\delta \omega) =  \frac{\lambda Q_{B}^{2}}{\pi^{3}} \int_{0}^{\pi/\omega} \left[ (\sin
\omega r )^{2} - (\sin (\omega + \delta \omega)r)^{2}\right]^{2} \frac{dr}{r^{2}} \ee 
Next, we evaluate $I(\delta \omega)$ using $\sin (\omega + \delta \omega)r \approx (\delta
\omega r) \cos \omega r + \sin \omega r$. Keeping terms to $\mathcal{O}(\delta \omega)$ gives  
         \be  I(\delta \omega) = \frac{4\lambda Q_{B}^{2} \delta \omega^{2} }{\pi^{3}}
\int_{0}^{\pi/\omega} \sin^{2} \omega r \cos^{2} \omega r dr \ee 
         Minimization of $\mathcal{E}_{\omega, \delta \omega}$ with respect to $\delta \omega$
yields 
         \be \delta \omega = - \frac{\pi^{2} \omega}{ \lambda Q_{B}} . \ee
For realistically large values of $Q_{B}$~\cite{Kusenko:1997si}, $\delta \omega$ is very small
compared to  $\omega$ which means $\omega_{L} \approx \omega_{B}$. We will neglect the difference
between $\omega_{L} $ and $ \omega_{B}$ in the remainder of the paper. 

Let us now evaluate the final state (off-flat-direction) Q-ball mass 
\be \label{massf} M_{f} = \omega_{f} (Q_{f}+Q_{B}) + \frac{4\pi U_{0}}{3 \omega_{f}^{3}} +
\lambda \int d^{3}x |L^{2} - q^{2}|^{2}   
\ee
Although this is a $(B+L)$-ball, we are interested only in allowing the sleptonic part to 
change its lepton number. Thus to ensure that $Q_{B}$ does not change as we vary $\tilde{L}$, we
demand that 
\be 
\frac{q_{i}^{2}}{\omega_{i}^{2}} = \frac{q_{f}^{2}}{\omega_{f}^{2}}.  
\ee
We recall that on the flat direction $L_{i} = q_{i}$, but this does not hold away from the flat
direction. To calculate the potential away from the flat direction, we use
\be |L^{2} - q^{2}|^{2} = \left(\frac{ \sin \omega r}{\omega r}\right)^{4} |L_{f}^{2} -
q_{f}^{2}|^{2}. 
\ee
Using  the definition of the lepton charge, eq.~(\ref{charges}), one can relate final and initial
leptonic amplitudes: 
\bea L_{f} &=& \left(\frac{Q_{f}}{Q_{i}}\right)^{1/2}
\left(\frac{\omega_{f}}{\omega_{i}}\right) L_{i} 
= 
\left( \frac{Q_{f}}{Q_{i}}\right)^{1/2} q_{f}
. 
\eea

Using the flat-direction expression for slepton amplitude (\ref{charges}) and rewriting in
terms of $N = Q_{i} - Q_{f}$, we obtain a simple approximate expression for the potential term: 
\be 
\lambda \int d^{3}x |L^{2} -q^{2}|^{2} = \frac{\lambda \alpha_{4} \omega_{f}}{\pi^{3}}
N^{2}, 
\ee
where $\alpha_{4} \approx 0.67$ is a numerical constant defined by 
\be
\alpha_{n} \equiv \int_{0}^{\pi} \left(\frac{\sin x}{x}\right)^{n} x^{2} dx. 
\ee
In particular, $\alpha_{4} =Si(2\pi) - \frac{1}{2} Si(4\pi) \approx 0.67$ and $\alpha_{6} \approx
0.39$. 
Using this expression, we can carry out the minimization of eq.~(\ref{massf}) and find that the
Lagrange multiplier $\omega$ is 
\bea \omega_{f} &=& \left( \frac{4 \pi^{7} U_{0}}{\pi^{3} (Q_{f}+Q_{B}) + \lambda \alpha_{4}
N^{2}}\right)^{1/4} = 
\omega_{i} \left(\frac{1}{1-N/Q_{i}}\right)^{1/4} \left(\frac{1}{1+{\lambda
\alpha_{4} N^{2} \left /\pi^{3} Q_{f} \right .  }}\right)^{1/4}
\eea
	Thus in the limit $Q_{i}$,$Q_{f} \gg N$, we see that $\omega_{f} \approx \omega_{i}$. We 
therefore take $\omega_{f} = \omega_{i}$ in the following.  We will verify the self-consistency of
this assumption \emph{ex post facto}. Using the value of $\omega_{f}$ away from the flat
direction, one can  compute the final state Q-ball mass as well as the number of neutrinos
emitted in the transition from flat-direction to off-flat-direction states. Since the final state
Q-ball is lighter than the original flat direction Q-ball the excess energy is released in
neutrinos:
	\be \Delta M \equiv M_{i} - M_{f} = \omega_{i} N - \frac{\lambda \alpha_{4}\omega_{i}}{\pi^{3}} N^{2} \ee
	
	For the emission to be energetically allowed, one must have $\Delta M \ge N m_{\nu}$.
This gives an upper bound on the number of neutrinos that can be emitted as the Q-ball transitions
from the initial flat direction state to a state with the VEV that is slightly off the flat
direction:
	\be N \le \left(\frac{\pi^{3}}{\lambda \alpha_{4}}\right) \left(1-
\frac{m_{\nu}}{\omega_{i}}\right). \ee
	
	For $\lambda \sim 0.1$ and an initial lepton number $Q_{i} =10^{24}$ and $\omega_{i}
\sim U_{0}^{1/4} Q_{i}^{-1/4} \sim 1$~MeV, as expected from
cosmology~\cite{Kusenko:1997si,Enqvist:2003gh,Dine:2003ax}, 
this gives the number of emitted neutrinos $N \lesssim 460$.  

We note in passing that $m_{\nu}$ must be less than the $\omega(Q) $ for emission to be possible.
The physical reason for it is clear, since $\omega$ can be thought of as the mass of a scalar
particle in the condensate.  In most models $\omega(Q) \gg
m_{\nu}$~\cite{Enqvist:2003gh,Dine:2003ax}.

The potential we have chosen in eq.~(\ref{massf}) is of the form one often
encounters for MSSM flat directions in gauge mediated models~\cite{Dvali:1997qv}. In particular, the
results of this section apply directly to the flat directions $QdL$ and 
$QQQL$~\cite{Gherghetta:1995dv}.

\section{Example: $Q_{1} \bar{d_{2}}L_{1}$ direction in the MSSM}
\label{neutral}

Let us now analyze a realistic model in which we consider Q-ball formation along the $Q_{1}
\bar{d_{2}}L_{1}$ direction.  This direction has been studied in detail by Dine, Randall and
Thomas~\cite{Dine:1995uk,Dine:1995kz} in the context of a gravity-mediated Affleck-Dine
baryogenesis. This direction is useful for our purposes since it carries nonzero baryon and lepton
numbers. It is useful in the AD scenario because it carries nonzero $B-L$ (in particular $B-L$ =
-1), which it must if one would like to have a nonzero baryon number after $B+L$ violating sphaleron
processes go out of equilibrium. This flat direction may be parameterized by the complex field
$\phi$ which is the VEV given to the squarks and slepton fields
\be Q_{1} = \frac{1}{\sqrt{3}} \left( \begin{array}{c}
\phi  \\
0 \end{array}  \right), ~~~
L_{1} = \frac{1}{\sqrt{3}} \left(\begin{array}{c}
0 \\
\phi \end{array} \right), ~~~
\bar{d_{2}} =\frac{1}{\sqrt{3}} \phi,  
\ee
where color indices are suppressed and subscripts label generation. The scalar potential on
the flat direction is identically zero in the supersymmetric limit, since the $F$-terms and
$D$-terms along this direction vanish. This degeneracy is lifted however, by soft SUSY breaking
terms, as well as by the higher-dimensional operators.   

The SU(2)$\times$U(1) D-terms\footnote{The SU(3) D-terms vanish because the two
squark fields, $Q$ and $d$, have the same amplitudes and the same value of $\omega$.}  are  
\bea \label{true} U_{D}   &=& \frac{g^{2}}{8}\left( |Q_{1}|^{2} - |L_{1}|^{2}\right)^{2} +
\frac{g'^{2}}{72}\! \! \left ( |Q_{1}|^{2} - 3|L_{1}|^{2} + 2
|\bar{d_{2}}|^{2}\right )^{2} , \eea
where $g = e/ \sin \theta_{w}$ is the SU(2) coupling and $g' = e/ \cos \theta_{\rm w}$ is the
U(1)$_{\rm Y}$ coupling.
When $Q_{1}^{1} = L_{1}^{2} = \bar{d_{2}} = \phi$ the D-terms in the potential vanish as they must.

The non-renormalizable superpotential along a flat direction has the form 
\be W_{NR} = \frac{\lambda}{nM^{n-3}}X^{k}  = \frac{\lambda}{nM^{n-3}}\phi^{n} , \ee 
where $n=mk \ge 4$ and $X=\phi^{m}$ is a gauge invariant composite operator of the fields that make
up the flat direction. For the case at hand $X=Q_{1} \bar{d_{2}}L_{1} = \phi^{3}$. $M$ is a large
mass scale usually taken to be either the GUT or the Planck scale. This gives a contribution to the
scalar potential 

\be V_{NR}(\phi) = \frac{|\lambda|^{2}}{M^{2n-6}}(\phi^{*} \phi)^{n-1}. \ee 

For the flat direction of interest, the lowest order contribution is for $n = 4$, so 

\be V_{NR}(\phi) = \frac{|\lambda|^{2}}{M^{2}} |\phi|^{6}.  \ee

The additional contribution to $\Delta M$ can affect the decay of the  flat direction
Q-ball. The new contribution is $\Delta M \supset \int d^{3}x (V_{NR}(\phi_{FD})
-V_{NR}(\phi_{G}))$, where $\phi_{G}$ and $\phi_{FD}$ are the field values in the ground state and
flat direction configurations, respectively.  Since both squark and slepton field amplitudes in
this case are different from their flat-direction values, they both make a contribution to the
non-renormalizable part of the potential:
\be 
(\Delta M)_{NR} = \frac{|\lambda|^{2}}{M^{2}} \int d^{3}x \left( |L_{i}|^{6} -
|L_{f}|^{6}\right) + \left( |q_{i}|^{6} - |q_{f}|^{6}\right) .
\ee

Using the values for $L_{f}$ and $Q_{f}$ found from the previous section, and assuming $N \ll
Q_{i}$, we obtain 
\be (\Delta M)_{NR} = \frac{-|\lambda|^{2} \alpha_{6} \omega_{i}^{3} Q_{i}^{2} N}{24\pi^{6} M^{2}}.
\ee
This translates into the following bound on $N$

\be N \le \frac{\pi^{3}}{\lambda \alpha_{4}} \left( 1- \frac{m_{\nu}}{\omega_{i}}  - \frac{|\lambda|^{2} \alpha_{6} \omega_{i}^{2} Q_{i}^{2} }{24\pi^{6} M^{2}} \right). \ee

Taking $\lambda \sim 0.1$, $M\sim 10^{18}$~GeV, we see that, with the inclusion of the
non-renormalizable term, the neutrino emission becomes energetically forbidden for $Q_{i}
\gtrsim 10^{25}$.

\section{Charged lepton emission for QdL}
\label{charged} 

In addition to the emission of neutrinos, there are, of course, other channels of decay that can
alter the Q-ball's lepton number, while leaving the baryon number unchanged.  The
transition from the flat direction Q-ball to the true ground state could be accomplished via
the charged lepton emission, for example, via emission of electrons, muons, and tau leptons.  As
before, the emission is only possible as long as $\omega$ is greater than the mass of the lepton. 
In most cases, the relic Q-balls have the value of $\omega $ below the muon 
mass~\cite{Enqvist:2003gh,Dine:2003ax}.  However, the emission of electrons remains a possibility. 
One expects the number of charged leptons emitted from the Q-ball to be smaller than the number of
emitted neutral leptons because the emission of a each electron adds electrostatic
energy to the final state Q-ball. The primary difference in this case is that the final state Q-ball
is electrically charged. Now the Q-ball mass with electric charge $eN_e$ is given by 
\be  
M_{f} = \omega_{f} (Q_{f}+Q_{B}) + \frac{4\pi U_{0}}{3 \omega_{f}^{3}} +\frac{3 e^{2}N_{e}^{2}
\omega_{f}}{20 \pi^{2}} + \lambda \int d^{3}x |L^{2} - q^{2}|^{2}.  
\ee 
Here we have used the expression for the electrostatic energy of a uniformly charged sphere and
$R(Q) = \pi /\omega(Q)$. As before we can find an upper bound on the number of emitted electrons
from $\Delta M \ge N m_{e} $. This yields
\be 
N \le \left( \frac{20 \pi^{3}}{3\pi e^{2} + 20 \lambda \alpha_{4}}\right) \left(1-
\frac{m_{e}}{\omega}\right). 
\ee
Taking $\lambda = 0.1$ and $\omega_{i} \gg m_{e}$, we obtain the bound $N \lesssim 280$ on the
number of electrons emitted when we consider only the electrically charged decay channel

As before, one must have $\omega(Q_{i}) \ge
m_{e}$ for emission to be kinematically allowed.  Since $\omega_{i} \sim U_{0}^{1/4} Q_{i}^{-1/4}$, 
this implies $Q_{i} \lesssim 10^{28}$. If this bound is not satisfied than electron emission is
energetically forbidden. We note that the same constraint for the neutrino case would allow much
larger values: $Q_{i} \lesssim 10^{56}$ (here we have taken $m_{\nu} \sim 0.1$~eV).

\section{Competing decay channels} 
\label{both}

\begin{figure}[htbp]
\centering
\epsfig{file=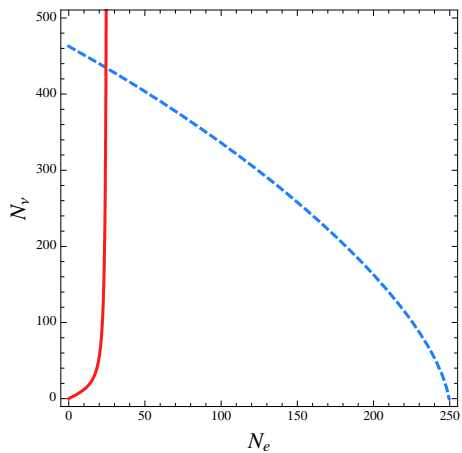, width=6cm}
\caption{ The region below  the dashed curve contains the allowed parameters for the decays
from FD Q-balls.  The solid curve shows the trajectory in the $N_{\nu}$-$N_{e}$ plane determined by
the respective decay rates into the electrons and neutrinos.  Here we show a representative
initial Q-ball charge, $Q = 10^{24}$. }
\label{fig1}
\end{figure}

The relic Q-balls can emit both neutrinos and electrons until they reach the true ground state
describe above.  In each particular case, the history of the AD condensate fragmentation and the
subsequent collisions and evaporations of Q-balls should be considered.  However, it is
instructive to consider a simplified history of Q-ball formation.  Let us assume that at some
initial time the Q-ball is in its flat-direction state.  This state would be the ground state of an infinitely large Q-ball because the mass per unit global charge becomes smaller than the mass of the lightest massive fermion in the limit $Q\rightarrow \infty$.  The AD condensate, from which the Q-balls form by fragmentation, can be thought of as a very large, superhorizon-size Q-ball.  Hence, it is reasonable to take the flat direction as the initial state of a Q-ball and to consider relaxation of this state into the ground state by emission of fermions, such as  neutrinos and electrons.  The relative rates of decay  determine the
final electric charge of the Q-ball.   Let us  find the lowest energy state achieved when both decay
channels are open. The Coulomb forces disfavor the electron emission. With both channels open, the
condition for decay $\Delta M \ge \sum_{i} N_{i} m_{i}$ becomes 
 \be \label{equal} \omega  \left[(N_{e} + N_{\nu}) - \frac{\alpha_{4}\lambda (N_{\nu}+ N_{e})^{4}}{
\pi^{3}} -\frac{3e^{2}N_{e}^{2}}{20  \pi^{2}} \right] \ge N_{e} m_{e} + N_{\nu}m_{\nu}. \ee  
With two lepton decay modes, the energetic constraint does not give us a fixed upper bound on
$N_{e}$, but instead merely carves out an energetically allowed region in the $N_{e}-N_{\nu}$ plane
(See Fig.~\ref{fig1}).  This region in interesting in that it allows the electron emission of a few
hundred.  Of course, such a larger emission is not realistic since the rate of electron emission is
expected to be much smaller than the rate of neutrino emission due to the Coulomb suppression.  It
is, therefore, the emission rates that determine the final ground state. 

For a sufficiently large Q-ball  the rate of decay into fermions is limited by the number of
fermionic states that can form inside and cross the boundary per unit time~\cite{Cohen:1986ct}.  
The maximal rate of fermion emission from the surface of a Q-ball 
\cite{Cohen:1986ct} is 
\be \frac{dQ}{dt} = -\frac{A \omega^{3}}{192 \pi^{2}}, \ee
where $A$ is the Q-ball surface area.  This rate is found by calculating the expectation value of
the normal component $ \langle {\bf n} \cdot  {\bf j} \rangle $ of the lepton current $ \bf j $ in
the frequency range $0 \le k \le \omega$, as in Ref.\cite{Cohen:1986ct}\footnote{Our upper 
limit is $\omega/2$, rather than $\omega$, since the model of Ref.~\cite{Cohen:1986ct} has a scalar
decaying into two fermions.
}:   
\be \langle {\bf n} \cdot  {\bf j} \rangle =  \frac{1}{(2\pi)^{3}}  \int_{0}^{k_{max}}  k^{2} dk
\int_{0}^{1} \cos \theta d \cos \theta \int_{0}^{2\pi} d\phi = \frac{\omega^{3}}{192 \pi^{2}} .
\ee
This calculation only applies to electrically neutral, massless neutrinos.   To extend this result
to massive, electrically charged leptons (electrons) one must take into account the Coulomb
interactions that change the upper bound on the frequency.  Let us denote this upper bound
$k_{max}$. For massive,
neutral particles $k_{max} = \sqrt{\omega^{2} - m^{2}}$.   Including the effect of Q-ball charging,
we obtain 
\be 
\omega = \sqrt{k_{max}^{2} + m^{2}} + \frac{3 e^{2} Z_{Q}^{2}\omega }{ 20 \pi^{2}}, 
\ee
where $Z_{Q}$ is the electric charge of the Q-ball (this is $N$ in previous notation).  For any
type of lepton emission the rate of fermion emission from the surface is 

\be  \label{rate} \frac{dQ}{dt} \lesssim\frac{A}{24\pi^{2}} k_{max}^{3}. \ee

We can use eq.~(\ref{rate}) to track the electron and neutrino emission, in order to
determine where in the $N_{e}-N_{\nu}$ plane a Q-ball will end.  We plot this trajectory in the
$N_{e}-N_{\nu}$ plane as the solid curve in Fig.~\ref{fig1}.  The intersection of the energetically
allowed boundary (dashed curve) and the dynamical path along which the Q-ball evolves (solid curve)
gives the maximal electric charge the Q-ball can acquire starting from a (electrically neutral) 
flat-direction Q-ball. This upper bound\footnote{In the massless limit we find only a slightly
different value for the number of emitted electrons $N_{e} \le \frac{\pi}{e} \sqrt{ 20/3}
\approx 27 $.} is $N_{e} \le 25$.

\pagebreak

\section{Conclusion} 

A large number of flat directions in MSSM and other supersymmetric models have both baryon number
and lepton number simultaneously.   Q-balls with VEVs along these flat directions can be entirely
stable: they owe their stability to the fact that a collection of protons and neutrons with the same
baryon number would have a larger mass than the Q-ball mass.  However, the true ground state of such
Q-balls can be different from the na\"{\i}ve ground state used in the literature.   In general,
the lepton component of a mixed baryoleptonic Q-ball deviates from the flat direction.   We have
shown this by starting with the na\"{\i}ve, flat-direction ground state and following its decay
into the true ground state plus some number of leptons.  The true ground state can be electrically
charged, which can have important ramifications for the experimental search for relic Q-balls.  

\section*{Acknowledgments}
We thank John M. Cornwall for very helpful discussions.  We also thank Antoine Calvez and Anupam
Mazumdar for helpful comments.  A.K. thanks Aspen Center for Physics for hospilality.  This work was
supported in part by the DOE grant DE-FG03-91ER40662 and by the NASA ATFP grant NNX08AL48G.

\bibliography{qball}
\bibliographystyle{h-physrev4}

\end{document}